\documentclass[twocolumn,showpacs,preprintnumbers,amsmath,amssymb,nofootinbib,prd]{revtex4}
\usepackage{epsfig}

\begin{document}

\title{Vortex Structures in a Rotating BEC Dark Matter Component}

\author{N.T. Zinner}
\affiliation{Department of Physics and Astronomy, Aarhus University, 
DK-8000 Aarhus C, Denmark}

\date{\today}

\begin{abstract}
We study the effects of a dark matter component that consists of bosonic particles with ultralight 
masses in the condensed state. We compare previous studies for both 
non-interacting condensates and with repulsive two-body terms and show consistency between 
the proposals. Furthermore, we explore the effects of rotation on a superfluid dark matter 
condensate, assuming that a vortex lattice forms as seen in ultracold atomic gas experiments. The 
influence of such a lattice in virialization of gravitationally bound structures and on 
galactic rotation velocity curves is explored. With fine-tuning of the bosonic particle mass 
and the two-body repulsive interaction strength, we find that one can have sub-structure on 
rotation curves that resembles some observations in spiral galaxies. 
This occurs when the dark matter 
halo has an array of hollow cylinders. This can cause oscillatory behavior
in the galactic rotation curves in similar fashion to the well-known effect of the spiral arms. We also consider how future
experiments and numerical simulations with ultracold atomic gases could tell us more about such exotic dark matter proposals.
\end{abstract}


\maketitle

\section{Introduction.}
In recent decades we have witnessed a breakthrough in precision cosmology with observational 
data from measurements of distant Supernovae \cite{riess98,perlmutter99} and from WMAP 
\cite{WMAP} indicating that our Universe is flat and dominated by non-standard 
forms of matter and energy. The data tell us that only about 4\% of the energy resides 
in the baryonic or luminous matter that we can observe through our telescopes. The rest of the energy is 
divided between the mysteries dark matter (26\%) and dark energy (70\%). The existence of dark 
matter (DM) was inferred earlier by its gravitational influence on the luminous matter in galaxies. 
Here the rotation velocity curves were found to disagree with expectations from general 
relativity. This 
lead people to introduce a dark matter halo around the galaxy that could explain 
the orbits. Models based on modification of the gravitational force are also being investigated 
but thus far such an explanation seems improbable \cite{zhytnikov94}.

The nature of this non-baryonic DM is still a source of heavy theoretical
and experimental investigation \cite{bergstrom2000}. 
Over the years numerous matter particles have been suggested as candidates for the invisible dark 
matter. The 
most popular ones are the weakly interaction massive particles (WIMP) which have very low 
reaction cross sections with normal baryonic matter \cite{overduin04}, although, since it is 
non-zero, 
there are many proposals and on-going efforts to measure them directly \cite{DMmeasurement}.
One idea is that DM consists
of scalar particles with extremely small masses, such as those appearing naturally as dilaton fields
\cite{dvali2001} in cosmological applications of string/M-theory \cite{kallosh2002}.
Ultralight DM was introduced in connection with a late-time
cosmological phase transition \cite{hill1989}, and assumed to be a Bose liquid in the condensate state \cite{sin94}.
Later it was shown that problems with small-scale structure
in the otherwise successful cold dark matter (CDM) model could be remedied by a Bose-Einstein condensate (BEC) DM component
\cite{hu00,goodman00,peebles00,schunck03}, since it will behave as non-relativistic CDM with a large
quantum coherence length, avoiding the cusp behavior of DM halos found in simulations \cite{hu00}.
In Ref.~\cite{sin94}, and more recently in Refs.~\cite{schunck1998} and \cite{bohmer07}, galactic rotation curves were 
calculated for BEC DM and compared to observations of dwarf and spiral galaxies. 

Here we address the effects of rotation under the assumption that the dark matter 
halos are Bose-Einstein condensates (BEC) of ultralight particles. When a BEC is rotated at a 
rate exceeding some critical frequency, quantized vortices can be formed \cite{BECbook}. The 
effects of 
vortex formation in galaxy halos has been addressed in Refs.~\cite{silverman02,mielczarek07}.
In Ref.~\cite{silverman02}, the vortex density of the Andromeda Galaxy M31 was considered, whereas 
Ref.~\cite{mielczarek07} discusses a vortex as the cause of the flattened galaxy rotation curve. There 
has also been some interest in analogues between rotating spacetimes and superfluid systems 
\cite{chap05}. The latter is a continuation of recent claims by Chapline and 
collaborators that Black Hole event horizons represent quantum phase transitions as seen in Bose 
fluids and therefore should not be regarded as a breakdown of classical general relativity 
\cite{chap03}.

Vortex lattices are seen in experiments with BEC \cite{yarm79,mad00,abo01,eng04} when the sample 
is rotated at rates above the 
critical frequency. In those systems one has a superfluid single-component BEC of alkali atoms at 
very low temperature with a two-body repulsive interaction described by the scattering length.
Recently it was shown how an array of lasers can actually induce gravity-like interactions 
by averaging over the usual electromagnetic dipole-dipole force \cite{odell2000}. 
This would in principle 
allow one to study the gravitational dynamics of BEC matter in the laboratory. Rotating such a 
system would allow us to study the properties of vortices when 
long-range $1/r$ attraction is present and perhaps reveal how structures are formed.
An alternative and likely more feasible way to approach the issue is through numerical simulations.
Extensive theoretical and experimental work in ultracold atomic gases has shown that Bose-Einstein
condensates are accurately described by the non-linear Gross-Pitaevskii equation \cite{BECbook}. Numerical 
simulations of systems governed by this equations including rotation and a gravitational potential
would therefore presumably yield information on structure formation and vortex dynamics.

In this article we want to explore some effects of rotating dark matter halos that 
consist of ultralight bosonic particles in the condensed state. This sort of dark matter would be 
superfluid and if rotated above the critical angular velocity it would allow the formation of a 
vortex lattices in the halo cloud. Section~\ref{non} considers the 
non-interacting case, whereas bosons with repulsive self-interactions are discussed in Section~\ref{int}. 
Effects of rotation on superfluid condensates are considered in Section~\ref{secrot} 
in the light of previous works addressing this possibility in connection with 
halos of spiral galaxies. In Section~\ref{secvort} we explore the influence of vortex lattices in 
dark matter halos for early structure formation, in the virialization of gravitationally bound 
structures, and on the galactic rotation velocity curves. We compare the rotation curves in our
ultralight BEC dark matter model to some observations of spiral galaxies and find 
that wiggles in the rotation curves can be accomodated in our model within certain paramater regimes. 
An ultralight BEC dark matter component can thus be an additional source of oscillatory behavior in the 
rotation curves, although the main source is presumably the baryonic matter in the spiral 
arms to which it is correlated \cite{beauvais1999,dicaire2008}.
Section~\ref{secfine} contains a 
discussion of some issues related to the fine-tuning of the mass of the bosonic particles and of 
the interaction strength of the repulsive two-body term. Some comparison between the presented 
astrophysical ideas and the current status of experiments and numerical simulations with 
ultracold atomic gases are given in 
Section~\ref{secatom}. Possible ways to test such exotic dark matter proposals in the 
laboratory or on a computer are suggested. Finally, Section~\ref{seccon} contains a short summary and
conclusions.

\section{Ultralight BEC Dark Matter}\label{non}
In this section we consider the possibility of a BEC of ultralight particles as the sole 
dark matter component. This was first suggested in Ref.~\cite{sin94} and further explored in Ref.~\cite{hu00} and 
Ref.~\cite{silverman02}. The idea is that ultralight particle $m\ll 1$ eV will have very large de 
Broglie wavelengths which means that quantum statistical effects are important and macroscopic 
coherent 
lumps of matter can emerge. As shown in Ref.~\cite{silverman02}, these light Bose particles will have 
a transition temperature to the condensed state that is of order $T_c\sim 2\cdot 10^{9}$ 
K~\footnote{This 
estimate of the critical temperature is based on comparing the thermal de Broglie wavelength 
\cite{BECbook} to 
the mean inter particle distance. For condensation to set in these two quantities must be 
similar, giving the relation $k_B T_c\sim hc(\rho/m)^{1/3}$ where $\rho$ is the mean density of 
the Bosonic matter \cite{silverman02}. A more detailed 
calculation of the transition temperature for a uniform Bose gas yields the expression $k_B T_c 
\sim \hbar^2 (\rho/m)^{2/3}/m$. Using the latter relation yields an even higher value for $T_c$ 
and does not alter the argument in the text that the Bose particles are expected to be in the 
condensed state at the present epoch.\label{tcnote}}, which is the 
expected temperature in the Universe after about 1 second. The fall of the temperature with the 
expansion means that an overwhelming majority of the Bose particles will be found in the 
condensate state. For the moment we will assume that the  particles 
are non-interacting (we will relax this assumption below) and therefore only gravity acts on the 
system. Following Ref.~\cite{hu00}, we now resort to Jeans instability analysis to estimate our 
parameters. The growing mode under gravity is $e^{\gamma t}$ with $\gamma^2=4\pi G\rho$, whereas 
the free field will be oscillatory; $e^{-iEt}$ with $E=k^2/2m$. The latter can be written as 
$e^{\gamma t}$ with $\gamma^2=-(k^2/2m)^2$. Noting that this is like normal Jeans 
analysis with sound speed $c_{s}^{2}=k/2m$ we have $\gamma^2=4\pi G\rho-(k^2/2m)^2$. Setting this 
to zero, we get the Jeans scale
\begin{eqnarray}
r_J&=&2\pi/k_J=\pi^{3/4}(G\rho)^{1/4}m^{-1/2}\nonumber\\
  &=&55m^{-1/2}_{22}(\rho/\rho_b)^{-1/4}(\Omega_m h^2)^{-1/4} \,\text{kpc},
\label{jeans}
\end{eqnarray}
where $m_{22}=m/10^{-22}$ eV, and the background density is $\rho_b=2.8\cdot 10^{11}\Omega_m h^2 
M_\odot$ Mpc$^{-3}$. Below the Jeans scale the perturbations will be stable and above it they will 
behave as ordinary CDM \cite{hu00}. The stability below the Jeans scale is guaranteed by the 
uncertainty principle. If the particles are confined further, their momenta will 
increase and oppose the gravitational contraction.

From the Jeans scale in Eq.~\eqref{jeans} we can see that the mass has to be extremely small for 
the current scenario to be responsible for galaxy-size structures at present. A typical galaxy has 
a matter density of about $200\rho_b$ and a radius of order $r\sim10$ kpc. Using $\Omega=0.3$ and 
$h=0.7$ in Eq.~\eqref{jeans}, we see that a mass of precisely order $m\sim10^{-22}$ eV is needed 
for the condensed particles. As noted in Ref.~\cite{hu00}, the observational evidence can accommodate 
even lighter fields $m\lesssim 10^{-33}$ eV in quintessence models. Ref.~\cite{hu00} 
proceeds with one-dimensional simulations and show that such ultralight Bose particles could 
actually help solve cusp problem in dark matter halos by suppressing small-scale linear. This was 
confirmed in more detail in Ref.~\cite{mat01}.

As mentioned in the introduction, the rotation curve of galaxies is an important feature that dark 
matter explains very well. In Ref.~\cite{silverman02} the ultralight BEC dark matter scenario was used 
to predict the rotation curve and the results were compared to observations from the Andromeda 
galaxy M31. The model was found to agree with the data if the boson mass was in the range $m\sim 
10^{-24}-10^{-23}$ eV, in rough agreement with the analysis in Ref.~\cite{hu00}.

\section{Including Self-Interaction}\label{int}
\noindent There has been a number of investigations into dark matter models with Bose particles 
that consider the (likely) more realistic case of bosons with repulsive self-interactions 
\cite{peebles00,goodman00}. The motivation for this was the problem of too much sub-galactic 
structure that was mentioned above. The typical case considered is that of a quartic 
self-interaction 
$\lambda\phi^4$ ($\lambda> 0$) 
which is the generic textbook example of an interacting field theory. From an atomic physics 
point of view 
the quartic case is also very interesting as the mean-field approximation yields the 
Gross-Pitaevskii equation (GPE) for the 
condensate state that has proven very useful in describing BEC experiments 
\cite{BECbook}.

Even more interesting is the fact that there is actually an exact solution for the case of a 
self-gravitating Bose gas with a repulsive quartic term \cite{goodman00,bohmer07}. This comes 
about since in this case 
the GPE can be recast into the Lane-Emden equation with polytropic index $n=1$ which has the 
analytic solution
\begin{equation}
\rho(r)=\rho_0 \frac{sin(r/r_0)}{r/r_0},\,\, r_0=\sqrt{K/2\pi G}.
\label{prof}
\end{equation}
$K$ comes from the equation of state: $p=K\rho^2$. This latter relation can be found by simply 
estimating the ground-state energy of a Bose condensate with repulsive self-interaction 
\cite{BECbook}. In atomic physics, one would typically relate the interaction strength (specified 
by $\lambda$ above) to the scattering length for two-body scattering at very low energy. In our 
current notation this yields $K=gn/2m^2$, $n$ being the particle density and 
$g=4\pi \hbar^2 a/m$ is called the interaction coupling constant \cite{BECbook}. The sound 
velocity in the gas is given by $c_{s}^{2}=gn/m$. The observed dark matter halos do not have the 
profile given by Eq.~\eqref{prof} but, as noted in Ref.~\cite{goodman00}, there could be 
non-zero momentum particles present that would allow a more realistic power-law behavior outside 
the core.

In this repulsive self-interaction scenario one can estimate the mass of the Bose particles by 
using the first zero of Eq.~\eqref{prof}, $R=\pi r_0$, which is assumed to be the halo radius. 
Combining all the formula given above yields
\begin{equation}
m=\left(\frac{\pi^2\hbar^2 a}{G R^2}\right)^{1/3}=6.73\cdot 10^{-2} a^{1/3} R^{-2/3}\,\text{eV},
\label{mass}
\end{equation}
where $a$ is measured in fermi and $R$ in kpc. Taking $R\sim10$ kpc 
and $a$ between 1 fm and 1 nm (typical atomic magnitudes \cite{BECbook}), we obtain 
masses in the range from $m\sim 10$ meV to 1 eV. We immediately notice that this is many orders 
of magnitude away from the non-interacting case of the last section. A mass of $m\sim10^{-22}$ eV 
would require an extremely low scattering length of $a\sim3.3\cdot 10^{-61}$ fm (keeping the same 
halo size). This is of course extremely small and borders on the non-interacting case which tells 
us that there is consistency with the results presented in the previous section.

At this point one might wonder if the critical temperature estimates given earlier will be altered 
significantly by the fact that the bosons now have a repulsive two-body interaction. In the 
context of the uniform Bose gas model, this has been addressed in Ref.~\cite{baym99} where is was shown 
that the change in the critical temperature obeys $\delta T_c/T_c\sim a n^{1/3}$. If we measure 
$a$ in fermi, $m$ in eV, and $\rho$ in $10^{-23}$ g/cm$^{-3}$ (typical halo density) then we get 
$\delta T_c/T_c\sim 2\cdot 10^{-10} a (\rho/m)^{1/3}$. We therefore see that the correction to 
$T_c$ is extremely small for both $m\sim 10^{-2}-1$ eV and for ultralight $m\sim 
10^{-22}$ eV (remembering that $a\propto m^3$ according to Eq.~\eqref{mass}). We therefore see that 
the estimate $T_c>2\cdot 10^9$ K given earlier will not be altered by the 
repulsive self-interaction.

The considerations in this and the previous section rely on Newtonian gravity, and
general relativistic effects could modify the results. A general relativistic 
description of the BEC dark matter scenario is discussed in Ref.~\cite{bohmer07}, particularly the 
effects on the rotational curves of galaxies. It was found that rotational curves of a number of 
galaxies could be well reproduced, although the Newtonian analysis already produces a tangential 
velocity of test particles at the halo radius of 365 km/s, consistent with observations. Note also 
that we are tacitly assuming that the ultralight Bose particles are non-relativistic which was 
shown to be a very good approximation in Ref.~\cite{sin94,silverman02}. We therefore restrict our discussion 
to Newtonian gravity here.

\section{The Effects of Rotation}\label{secrot}
Several studies have addressed the effects of rotation on BEC dark matter. In Ref.~\cite{bohmer07} 
there is a brief discussion of the effect on the Lane-Emden equation, whereas Ref.~\cite{mielczarek07} 
consider a BEC of axions with a single vortex arising from global rotation in the early Universe. 
This latter scenario is, however, less likely to occur since the global rotation rate of the 
Universe 
can be estimated from various observations and is likely very small if non-zero \cite{CMBrot}. The formation of vortices in 
condensates is known from atomic physics \cite{BECbook}. For superfluid Bose systems any angular 
momentum imparted will reside in quantized vortices that constitute small regions where the 
superfluid density goes to zero on a scale comparable to the coherence length (to be discussed 
below). As already noted in 
Ref.~\cite{goodman00} BEC dark matter with self-interactions will actually constitute a 
superfluid.

A superfluid will only form these vortices when the rotation rate exceed a critical frequency 
$\Omega_{c}$. For a sample of size $R$ with coherence length $\xi$ the critical value is 
\cite{BECbook}
\begin{equation}
\Omega_c=\frac{\hbar}{mR^2}\ln\left(\frac{1.46R}{\xi}\right)=\frac{6.21\cdot10^{-17}}{m_{22}R^2}
\ln\left(\frac{1.46R}{\xi}\right),
\label{crit}
\end{equation}
where $m_{22}=m/10^{-22}$ eV and $R$ is measured in kpc. This critical frequency is based on the 
Gross-Pitaevskii equation and the coherence length $\xi$ 
is given by the self-interaction through $\xi=1/\sqrt{8\pi a n}$ with $n$ the particle density and 
$a$ the scattering length. The formation of vortices has been 
experimentally demonstrated in superfluid $^{4}$He \cite{yarm79}. Later they were produced in 
dilute BEC alkali atom gases \cite{mad00,abo01,eng04} where beautiful 
lattices where created with more than 100 vortices in regular alignment. These experiments also 
showed a density profile that had dips at the vortex positions, confirming the theoretical 
expectation of density drops in the vortex core mentioned above, although the actual experiments 
had cores with small but non-zero density. The expected increase in the number of vortices with 
increasing rotation frequency was also observed. Theoretically, the vortex line density is 
given by $n_v=m\Omega/\pi\hbar$ which yields a total number of vortices $N_v=\pi R^2 n_v$.

The idea of similar vortex formation in rotated BEC dark matter was first considered in 
Ref.~\cite{silverman02}. Here the rotation rate of the Andromeda galaxy M31 was calculated and compared 
to the critical frequency. M31 was found to rotate at a rate vastly above $\Omega_c$ and 
vortices were therefore suggested as a possible consequence. The total number of vortices for M31 
was 
found to be around $N_v\sim340$. As a possible means of detecting these vortices, 
Ref.~\cite{silverman02} suggested gravitational lensing or polarization effects from frame-dragging of 
light from distant background sources. The use of gravitational lensing to infer the possible 
existence of a BEC dark matter component was also suggested and explored in Ref.~\cite{bohmer07}.

As already mentioned above, the halo radius can be taken as the first zero of the solution in Eq.~\eqref{prof}, 
i.e. $R^2=\pi^2 \hbar^2 a/G m^3$. 
In the scalar field treatment with interaction $\lambda\phi^4$, the corresponding size estimate is 
$R^2=\pi^2 \lambda \hbar^3 / 8\pi G m^4 c$ \cite{arbey2003,schunck03}. Equating these, we obtain
$\lambda=8\pi a m c/\hbar=1.3\cdot 10^{-29} (a/1\,\textrm{fm})(m/10^{-22}\,\textrm{eV})$. 
Using $m=10^{-22}$ eV, a halo mass density 
$\rho_{\textrm{halo}}\sim 10^{-23}$ g/cm$^{-3}$, and assuming that $\xi\sim 1$ kpc, we find
$\lambda\sim 10^{-92}$, a very small value. As noted earlier, the repulsive self-interaction 
is not likely to influence 
the gravitational formation of the halo \cite{schunck03}. However, as discussed above such 
small values give large vortex cores when $\Omega\gg \Omega_c$. In Ref.~\cite{arbey2003}, 
fits to rotation curves were presented and it was found that 
$m^4/\lambda\sim 50-75$ eV$^4$. For $m=10^{-22}$ eV this gives $\lambda\sim 10^{-90}$, 
close to the present estimate. Importantly, our smaller $\lambda$ actually fulfills 
the Big Bang Nucleosynthesis (BBN) bounds discussed in Ref.~\cite{arbey2003}.

\section{A Dark Matter Vortex Lattice}\label{secvort}
If dark matter contains a component of condensed BEC particles that is superfluid and if 
the halos are rotating then it is not inconceivable that there can be vortex formation as 
discussed above. However, the quantized vortex discussion of Ref.~\cite{silverman02} makes an important 
assumption about the 
coherence length, $\xi$, entering $\Omega_c$ in Eq.~\eqref{crit}. $\xi$ is taken to 
be of kpc size. This is based on the non-interaction arguments of Section~\ref{non}. With no 
self-interaction there is only the gravitational scale $\hbar^2/GMm^2$ available, which becomes of 
galactic size for masses $m\sim 10^{-22}$ eV. However, when including self-interactions as in 
Section~\ref{int} through a scattering length $a$, there is also a scale given by 
$\xi=1/\sqrt{8\pi a n}$, which is the usual Gross-Pitaevskii coherence length. The latter 
coincides 
with the characteristic length over which the density is expect to go to zero in a vortex.

In the following discussion we will make the assumption that it is actually the coherence 
length $\xi$ that determines the vortex core size. This is reasonable 
since the vortices are local entities and as such should depend on the local 
interactions. From this point of view the gravitational scale $\hbar^2/GMm^2$ merely serves to 
determine the total halo cloud size. The coherence length can be written as
\begin{equation}
\xi=2.73\cdot10^{-32}m_{22}^{1/2}a^{-1/2}\rho_{DM}^{-1/2}\,\text{kpc},
\label{cohere}
\end{equation}
where $\rho_{DM}$ is the dark matter halo mass density measured in $10^{-23}$ g/cm$^3$, $a$ is in 
Fermi, and $m_{22}=m/10^{-22}$ eV. The measure for $\rho_{DM}$ originates from assuming that a 
halo has a radius of about 10 kpc and contains about $3\cdot 10^{11}$ M$_\odot$ of mass giving a 
density of about $1.74\cdot 10^{-23}$ g/cm$^{3}$. The front factor in 
Eq.~\eqref{cohere} seems exceedingly small and it would seem that any vortices on this scale would 
be 
completely irrelevant to the galaxy structure. However, we recall from the mass estimate in
Eq.~\eqref{mass} that $a$ also needs to be extremely small. Inserting a value of $a\sim10^{-61}$ fm 
(corresponding to $m\sim10^{-22}$ eV) actually gives $\xi\sim0.05$ kpc, so that the vortex size is 
sub galactic but sizable. This demonstrates that vortex structures could be important for the 
structure of rotating galaxies and we will discuss possible effects below.

Before we consider vortex lattice effects on structure and evolution, we now 
briefly compare our suggestions with the observational data from M31. 
Ref.~\cite{silverman02} estimate the critical frequency to be about 
$\Omega_c=2\cdot10^{-19}$ rad/s for M31 (using $R\sim150$ kpc and $m\sim2\cdot10^{-23}$ eV). The 
estimate 
for the actual rotation frequency of M31 based on observations is given as $\Omega=5\cdot10^{-17}$ 
rad/s, 
so that the assumption $\Omega\gg\Omega_c$ is justified and a vortex lattice is 
therefore possible. However, in Ref.~\cite{silverman02} the coherence length used in 
Eq.~\eqref{crit} was taken to be $\xi\sim30$ kpc (the gravitational length scale for the 
non-interacting condensate discussed above). We are assuming that the coherence 
length scale for vortex formation is that associated with the two-body repulsion. 
This is, however, not a serious obstruction since it enters only in the logarithmic term in 
Eq.~\eqref{crit}. Using the value quoted above of $\xi\sim0.05$ kpc in the critical frequency would 
therefore only amount to about a factor of five increase which would not jeopardize the relation 
$\Omega\gg\Omega_c$. Furthermore, changing the two-body repulsion (through changes 
in $a$) would not be severe for the same reason.

\subsection{Influence on Early Structure Formation}
For the vortex lattice to appear in the current scenario with BEC dark matter we need rotation.
One can assume that rotation is a primordial feature 
of the Universe. This possibility has been explored by many authors \cite{uni1,uni2,uni3,uni4}. However, 
this is at odds with considerations of CMB anisotropies \cite{silk70}, which suggest that the 
primordial rotation rate is very small \cite{CMBrot}. The rotation that one finds in galaxies is 
therefore not considered to be primordial and is believed to arise from gravitational tidal torque 
forces during the growth of non-rotating initial perturbations toward virialized galactic 
structures. This way of generating angular momentum is well studied \cite{angmon1,angmon2,angmon3} and widely 
accepted as a key mechanism.

Although primordial rotation seems out of the question, we will briefly 
consider what would be the result if it was found that our Universe could have been rotating in 
early epochs. More precisely we want to explore what would be the effect of a rotation rate large 
enough to support a vortex lattice at the time of decoupling when baryonic perturbations can start 
to grow. We can argue in the usual way that when baryonic perturbations can finally grow, they 
will do so under the influence of a potential well that is created by the dark matter 
perturbations. Assuming that the baryons themselves provide a negligible contribution to the 
background density, a Jeans analysis tells us that the baryonic density contrast in the linear 
regime obeys \cite{padma}
\begin{equation}
\ddot{\delta}_{B}+\frac{4}{3t}\dot{\delta}_{B}+\left(\frac{k_B 
T}{m_p}\right)\frac{k^2}{a^3}\delta_{B}=\frac{2}{3}\frac{\delta_{DM}}{t^2},
\label{linbary}
\end{equation}
where $t$ is cosmic time, $a$ is the scale factor, $T$ the temperature of the 
baryons, $m_p$ the proton mass, $k_B$ the Boltzmann constant, and $k$ the wave vector of the 
Fourier density mode under consideration. For a matter-dominated 
Universe with $\delta_{DM}\propto a$, one finds
\begin{equation}
\delta_{B}(t)=\frac{\delta_{DM}(t)}{1+Ak^2},
\end{equation}
where $A=3/2(k_B T_0/m_p)(t^2/a^3)$ with $T_0$ the temperature today (using matter-dominance from 
decoupling to the present). Baryons are thus coupled tightly to dark-matter on large scales, 
whereas the pressure support on small scales ($Ak^2\ll 1$) suppresses growth. If there was a 
primordial rotation of the dark matter of sufficient magnitude to 
cause vortex formation, then one would have to modify the driving term $\delta_{DM}$ on the 
right-hand side of Eq.~\eqref{linbary} to reflect the vortex structure. Given some characteristic 
vortex size (as in Eq.~\eqref{cohere}) one would have to make the Fourier transform of a 
swiss-cheese like configuration of the dark-matter. This would then give a new driving term and 
thus influence the structure formation in the baryons. There is a technical point here that we 
have ignored, which is that the Jeans analysis does not strictly apply to rotating systems. There 
would be corrections to this from Coriolis and centrifugal terms but we assume that these can (at 
least locally) be neglected. This assumption would of course have to be checked in more detailed 
investigations.

Since there is no evidence for primordial rotation, the discussion above is likely of little 
relevance to the Universe we live in. The rotation that we observe in spirals must therefore 
originate in late-time events when the perturbations have entered the non-linear regime and Jeans 
analysis is no longer useful. Estimates show that the galactic structures that we see today cannot 
not have formed at redshifts much larger than $z\sim 10$ \cite{longair}. Linear perturbation 
theory 
is expected to hold around decoupling which is $z\sim1000$, so we are well beyond this 
approximation. The structure of baryons is also complicated by pressure and radiative terms, which 
means that hydrodynamical codes must be used. State-of-the-art in structure formation are highly 
involved $N$-body computer simulations that can calculate the evolution of the initial spectrum of 
perturbations into the non-linear regime. However, these simulations do show that the 
gravitationally bound systems that arise tend to virialize on a fairly short timescale. In this 
article we will not consider such advanced simulation techniques but merely consider what the 
effects of vortices could be on the virialized structures, since these are the ones that we 
observe to be rotating.

\subsection{Influence on Virialization}
Let us now address the question of the virialization process itself. The usual 
argument is that the bound system collapses under gravity with internal heating as a result. The 
virial theorem tells us that the internal kinetic energy should be half the gravitational 
potential. The systems therefore collapses until a radius is reached where this condition is 
fulfilled. However, when we introduce short-range two-body repulsion for the BEC dark matter,
the virial theorem is modified and becomes
\begin{equation}
2E_{kin}+E_{grav}+3E_{rep}=0,
\end{equation}
where $E_{kin}$ is the kinetic energy, $E_{grav}$ the gravitational potential energy, and 
$E_{rep}$ is the energy arising from the two-body term. The latter contribution is easy to 
calculate for our zero-range potential and becomes
\begin{equation}
E_{rep}=\frac{2\pi \hbar^2 a}{m}\int d\vec{r} n^2,
\end{equation}
where $n(\vec{r})$ is the density of BEC dark matter. $N$-body simulations and 
observations tell us that objects virialize. So the contributions from $E_{rep}$ must be 
negligible in virialization to not jeopardize this fact. Consider the ratio
\begin{equation}
\frac{E_{rep}}{E_{grav}}=-\frac{5}{2} \frac{a\hbar^2}{G R^3 m^3}.
\label{repgrav}
\end{equation}
Here we are 
assuming a uniform sphere for the matter distribution which gives $E_{grav}=-3/5GM^2/R$. 
In order for $E_{rep}$ to not 
influence virialization, we need this ratio to be much smaller than 1. However, since we have the 
relation between $a$ and $m$ in Eq.~\eqref{mass} this ratio becomes simply $-5/2\pi^2\sim -0.25$, 
so 
we see that the repulsive interaction will be non-negligible. We therefore have to reconsider how 
to apply the virial theorem. This we will do along the lines of the simple top-hat model as 
discussed in Ref.~\cite{longair}.

The virial theorem with inclusion of the repulsive two-body interaction and the use of 
Eq.~\eqref{mass} for $m$ yields
\begin{eqnarray}
E_{kin}&=&-\frac{1}{2}\left(+E_{grav}+3E_{rep}\right)=
-\left(\frac{1}{2}-\frac{15}{4\pi^2}\right)E_{grav}\nonumber\\
&\approx&-0.12E_{grav}.
\end{eqnarray}
We see that the internal kinetic energy is a considerably smaller fraction of the gravitational 
energy than for standard virial consideration with $a=0$. If we have a sphere of 
matter with radius $r_{max}$ that starts to collapse from rest, then from conservation of energy 
one can easily estimate that the virial theorem will be satisfied when $r\approx r_{max}/1.14$, 
whereas 
for $a=0$ is would be $r=r_{max}/2$ \cite{longair}.
Virialization thus predicts a less compact system in the $a\neq 0$ case. This means that when 
gravitational objects become virialized they will have densities that are about $27\rho_b$,
as compared to the $a=0$ estimate of $150\rho_b$ 
\cite{longair}. The two-body repulsive interactions are thus seen to yield less dense objects, 
something that could be very good in terms of getting less dense halos than some simulations 
produce and that are at odds with observations. To test whether this helps would require full 
simulations with ultralight BEC dark matter.

$N$-body simulations predict that virialization occurs at densities above
$150\rho_b$ (estimates put the value close to $400\rho_b$ \cite{longair}), which is an 
order of magnitude larger than the calculation above.
Ultralight BEC dark matter would therefore seem to be ruled out.
However, the above results are only as good as the assumptions we use in their derivation. 
In particular, the use of Eq.~\eqref{mass} is based on the Lane-Emden result of Eq.~\eqref{prof}. We 
know that halos do not have this profile. We 
should therefore not put to much emphasis on the mass $m$ in Eq.~\eqref{mass}, but rather leave 
$m$ as a parameter. If we assume that $E_{rep}=-\kappa E_{grav}$, one can invert the virialization 
analysis of Ref.~\cite{longair} of to get an estimate of $\kappa$ if we are to get virialized densities 
of $\gtrsim 400\rho_b$. This yields $\kappa\lesssim 0.1$. For this to be fulfilled,  
Eq.~\eqref{repgrav} tells us that $a/m_{22}^{3}\lesssim  4\cdot 10^{-44}$ fm/eV$^3$ with $a$ in 
fermi and $m_{22}=m/10^{-22}$ eV. This shows that we can actually accommodate
$N$-body results on virialization are fulfilled by having a small value of $a$ for ultralight 
Bose particles. Such small values for $a$ are also consistent with the values predicted from the 
arguments in earlier sections of $a\sim 10^{-61}$ fm. 

\subsection{Influence on Galactic Rotation Curves}
Another interesting issue is whether a dark matter vortex lattice can influence the galactic 
rotation curve. These curves are key pieces of evidence for the existence of dark matter due to 
their flat (or slightly increasing) behavior at large radii where standard Newtonian theory 
including only luminous matter predicts a sharp decrease with radius. 
The galactic rotation velocity is calculated from the basic formula \cite{padma}
\begin{equation}
v^{2}(r)=\frac{GM(r)}{r}=\frac{4\pi G}{r}\int_{0}^{r}dx x^2 \rho(x),
\end{equation}
where $\rho(r)$ is the total mass contribution from dark and luminous matter. In the present 
work we assume that the dark halo contribution has the form
\begin{equation}
\rho_h(r)=\frac{\rho_0}{1+\left(\frac{r}{r_0}\right)^\gamma},
\label{halo}
\end{equation}
where $r_0$ is related to the core radius and $\gamma$ is the exponent determining the large distance
behavior. This form avoids the small distance cusp that BEC DM presumably solves \cite{hu00,goodman00,peebles00,schunck03}.
The rotation curve can now be 
calculated from the simple formula $v^2=v_{\textrm{dark}}^2+v_{\textrm{lum}}^2$, where $v_{\textrm{lum}}$is the
contribution from the luminous matter in the bulge, disk, and surrounding gas.

Experiments with rotating atomic BEC produce triangular vortex lattices. Theoretically, this is is the minimum
energy configuration, with the square lattice slightly higher in energy \cite{tkachenko1966}. Recent studies including
long-range dipole-dipole interactions between the bosons
find that other configurations are preferred as the long-range forces increase \cite{dipole2005-1}, and we speculate
that similar results can arise for the $1/r$ gravitational force. However, since it is unclear which kind of lattice would arise in the galactic halos, we have
explored both triangular and square lattices. 
We also mention that square lattices are commonly seen in two-component condensates \cite{schweikhard2004}, which would be relevant
for multiple species BEC DM.

\begin{figure*}
\epsfig{file=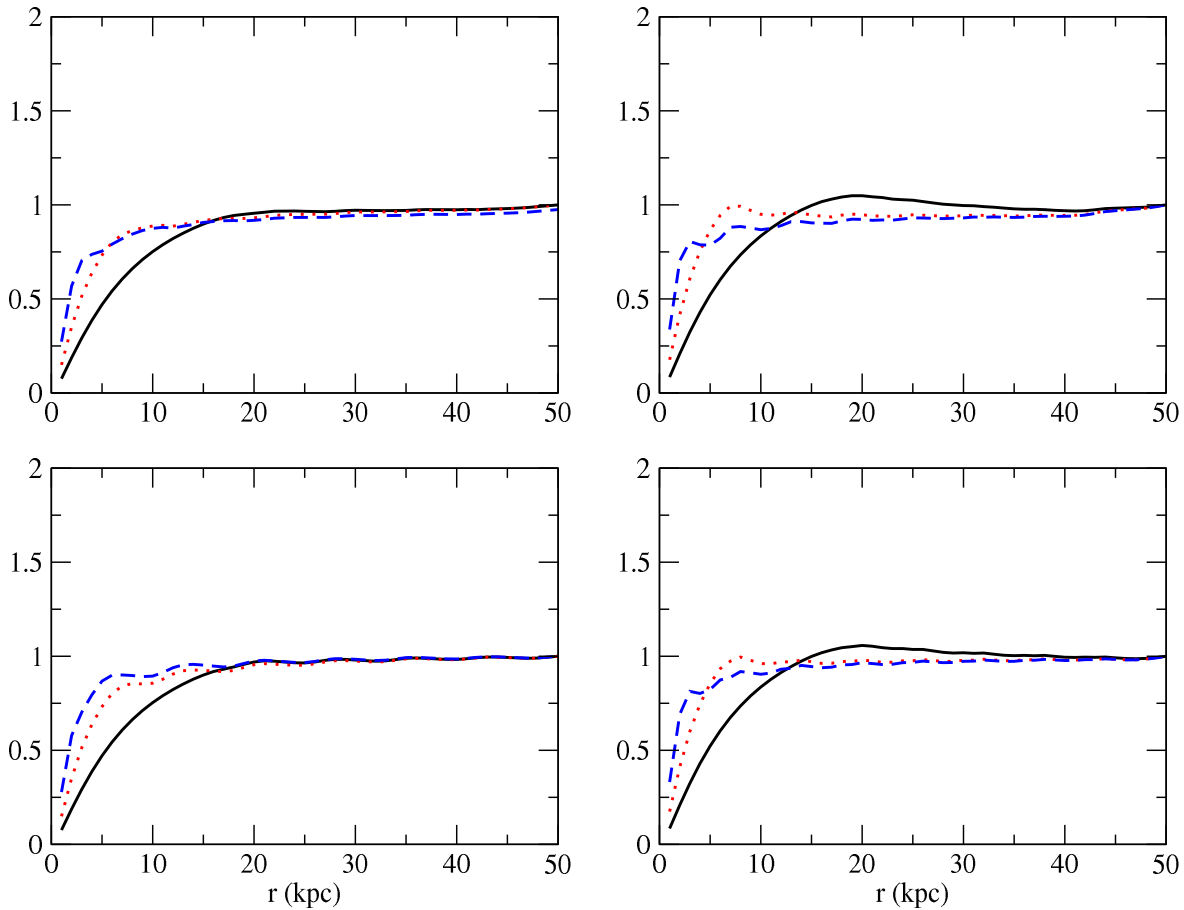,angle=270,clip=true,scale=0.6}
\caption{Rotation curves based on BEC DM with a vortex lattice for different halo sizes $r_0=8$ (full), $r_0=4$ (dotted), and $r_0=2$ kpc (dashed).
The vortex core size is 5 kpc. 
In the left column
the distance between vortices is 2.5 kpc, whereas it is 1 kpc in the right column. Upper row is for a triangular lattice and the lower
one is for a square lattice.} 
\label{testfig}
\end{figure*}

The vortex lattice is modeled as an array of zero-density tubular regions of square cross section in the DM mass density.
We have thus ignored the
spherical shape and the smooth fall-off to zero density that the more realistic vortex cores would have. We have checked that this
simplification did not have significant effects on our results. Due to this implementation, the vortex size represents
an effective coherence length, simply related to $\xi$. Keep in mind that the square shape can
cause some rather sharp features in the calculated rotation curves which are smoothed for more realistic
profiles. To find the rotation velocity we have numerically integrated this 'swiss-cheese' halo along with 
the luminous contributions. In effect the method used assumes that the lattice rotates with the system, which is what is 
seen in cold gas experiments as well. To demonstrate the effect of the vortex lattice, we show in Figure~\ref{testfig} a number of 
rotation curves based solely on the DM. These have been calculated for vortex size 5 kpc and vortex-vortex distances of 1 and 2.5 kpc. These
parameters are optimal in order to get the oscillations at short distances shown in Figure~\ref{testfig}. For smaller
core sizes the effect vanishes and for larger ones the scale of wiggles is too large to fit observations.

At this point we mention
that we have used a cutoff on the vortex lattice for small radii (of order the halo size $r_0$), effectively ignoring vortices that are located in the 
halo center. With this cutoff we can fit the observation of large constant density in the center of the galaxies. This appears reasonable since luminous matter should be dominant in the inner-most region. Another issue is the density of vortices. In our simple model we have assumed this to be constant with the use of regular lattices but there could be variations \cite{yu2002}.

The wiggles found in the velocity curves with vortices present are quite interesting in relation 
to observations. Here one also sees distinct wiggling features in the plots \cite{bohmer07,sin94}. 
In Ref.~\cite{sin94} the data from NGC2998 was explained by postulating the BEC dark matter to be in an 
excited condensate state which has several zeros in the wave function. When this is translated into 
a dark matter distribution it gives areas of low density in the halo. In the velocity curves this 
will of course have similar effects as the vortex lattice with its empty cores. However, with 
vortices one can get wiggles without having to explain how the excitation occurred and why the 
condensate should be in a particular excited state at present. 
One would of course also have to include
the luminous matter for a real comparison to data, and for small radii this will presumably dominate the rotation curves 
\cite{salucci2007}. Below we will include the effects of the luminous component explicitly when comparing 
to observations.

Before embarking on a comparison to observations, we note a number of problems with this model. 
First, we have assumed that 
the vortex cores have the same shape at all radii, independent of the local halo distribution. 
This is likely unrealistic as we would expect some effects of the finite size toward the edge. 
Secondly, for the coherence length to approach 5-10 kpc, from Eq.~\eqref{cohere}, we would need 
either $a\sim10^{-65}$ fm or an increase in $m_{22}$ by four orders of magnitude (or a combination 
of both). We estimated earlier that existing suggestions for the parameters of ultralight BEC dark 
matter of $a\sim10^{-61}$ fm and $m\sim10^{-22}$ eV gives $\xi\sim 0.05$ kpc, which would be 
impossible to see in velocity curves. So it would seem that we need some fine-tuning of $m$ and 
$a$ to get any effect. This we will address below.

\subsection{Comparison to Spiral Galaxies}
An interesting feature of the BEC DM model is the oscillations in the rotation velocity at small radii, and we now compare the model to spirals that show pronounced oscillatory features. Both dark and luminous matter play important roles in modeling spirals, and we therefore 
include the luminous component based on recommended values from various observation as described below. Note that we allow the dark
matter component and the details of the vortices to vary in order to best describe observations. For simplicity, we ignore
the feedback effect on the luminous matter of changing the dark component. Since luminous matter dominate the rotation curve
at small radii and dark matter at larger radii in our model, this should be a fair approximation.

\begin{figure*}
\begin{center}
\epsfig{file=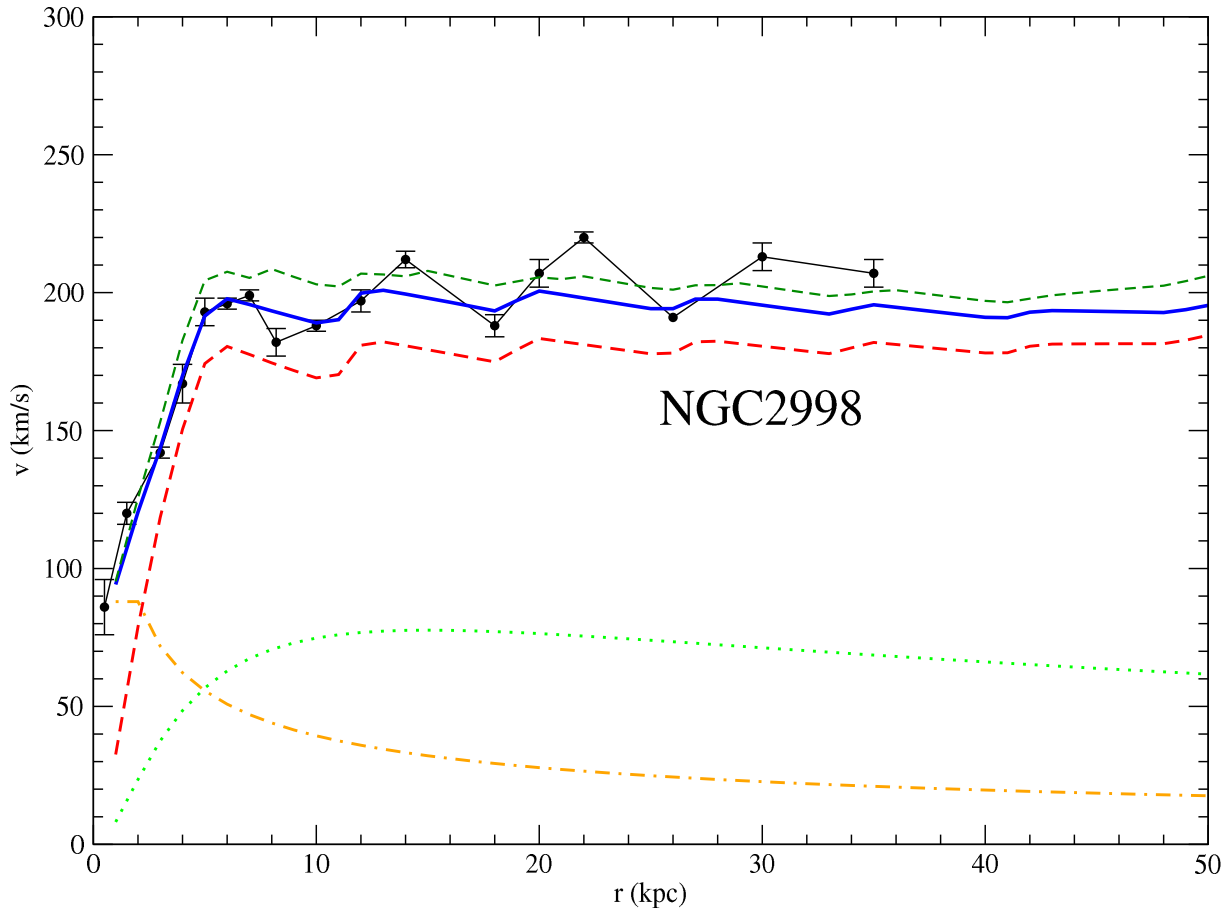,angle=270,scale=0.6,clip=true}
\caption{Rotation curve data for the Spiral galaxy NGC2998 (dots with uncertainties) and calculation with a 
square (thick line) and triangular (short dashed) BEC vortex lattice in the DM component. Shown are also the 
individual contributions from the bulge (dot-dashed), disk+gas (dotted), and DM for the square lattice case (long dashed).
The parameters used are $r_0=5.0$, $\gamma=2.15$, $b=3.0$, $L_\textrm{v}=6.5$, and $d=1.0$ (all lengths in kpc).} 
\label{2998fig}
\end{center} 
\end{figure*}

We consider a sample of spirals that have been used in BEC dark matter studies previously \cite{sin94,schunck1998,bohmer07}. In Figure~\ref{2998fig} we show data for the Sc spiral NGC2998 \cite{rubin1985}. The bulge was
assumed to give a constant contribution for $r<3$ kpc and then fall as $r^{-1/2}$ \cite{kent1986}, whereas the disk and gas are combined into the functional form $(r/(r+b))^6$ of range $b$ \cite{bohmer07}. The magnitudes of these non-dark contributions was scaled to reproduce those in Ref.~\cite{kent1986}, whereas the
DM contribution is varied to best describe the observations.
It can be seen 
in Figure~\ref{2998fig} that a square lattice is closer to the data. In particular, we see good agreement between maxima and minima of the curves, although the magnitudes are not well reproduced, particularly at larger radii. We find that the oscillations in the model are 
directly related to the vortex core size, $L_\textrm{v}$, and lattice distance, $d$, and the characteristic length of the 
wiggles observed in the rotation curves can thus be matched in the present model by careful choice of parameters. 
However, as explained before the main source of the oscillatory behavior is considered to be baryonic through due to the presence
of spiral arms.

\begin{figure*}
\begin{center}
\epsfig{file=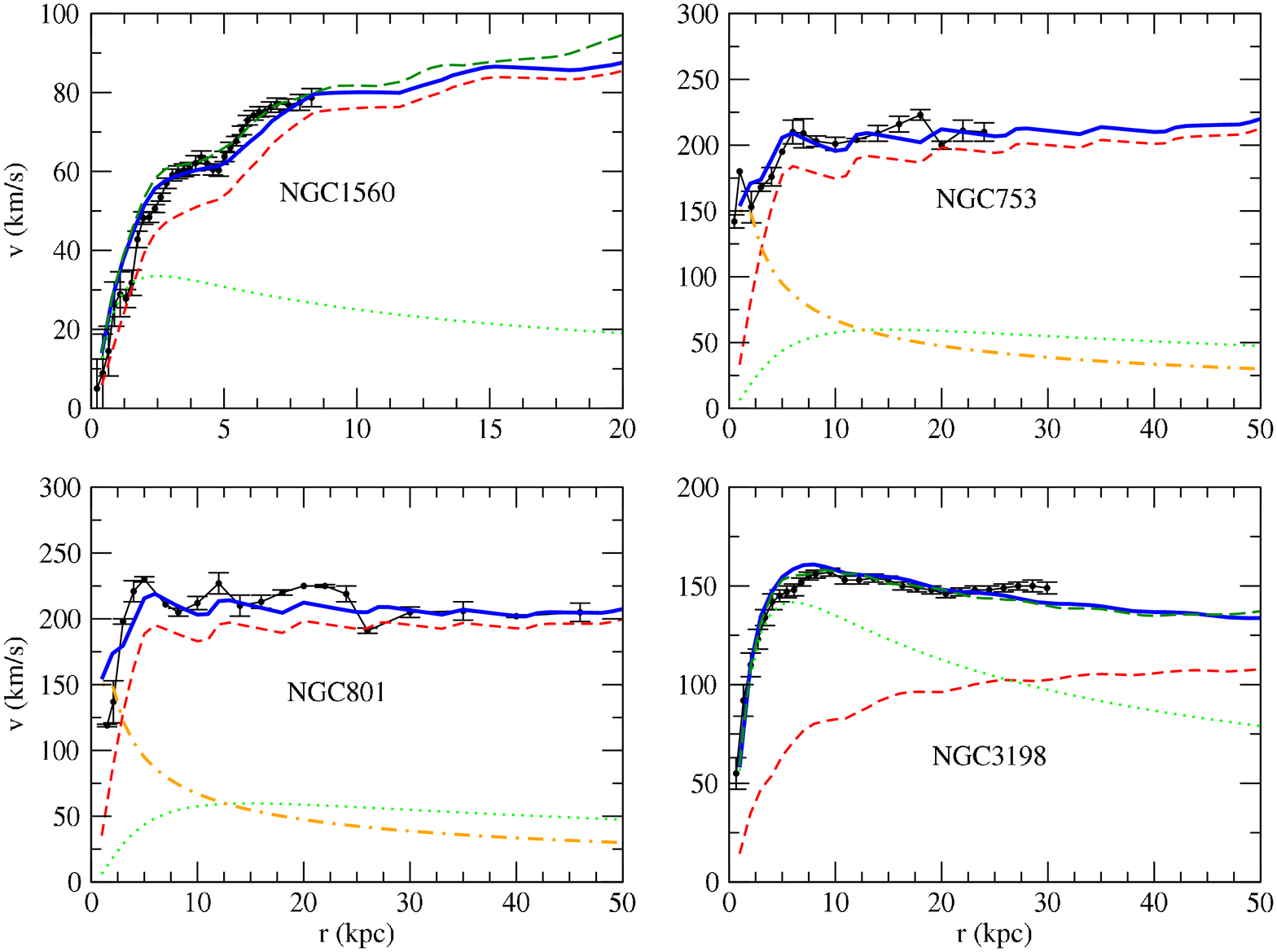,angle=270,clip=true,scale=0.6}
\caption{Rotation curves for NGC1560 (5.0,2.15,0.5,5.0,1.0), NGC753 (5.0,2.00,3.0,6.5,1.0), NGC801 (5.0,2.15,3.0,6.5,1.0), 
and NGC3198 (5.0,2.00,1.2,6.5,2.0) compared to model calculations. The plot assignments are
as in Figure~\ref{2998fig}. The parenthesis give the parameters used in the format $(r_0,\gamma,b,L_\textrm{v},d))$.
Notice the change of radial distance scale for NGC1560.} 
\label{gfig}
\end{center} 
\end{figure*}

To further explore the model, we consider observations from two Sc spirals, NGC753 and NGC801 \cite{rubin1985}, a dwarf spiral, NGC1560 \cite{broeils1992}, and a thin disk spiral, NGC3198 \cite{begeman1989} in Figure~\ref{gfig}.
The non-dark components are modeled as for NGC2998 above. Notice that 
NGC1560 and NGC3198 have no bulge. The DM for NGC753 and NGC801 assumes a square vortex lattice (a triangular lattice compares worse with data), and we see that this gives a better agreement with the observed wiggles than for NGC2998, but still with discrepancies between minima and maxima. This could likely be resolved, if one makes the 
vortex density non-uniform as suggested in Ref.~\cite{yu2002}. In the dwarf case of NGC1560, we see that our model can reproduce the observed kink in the curve, in this case even better with the triangular lattice. This feature was not reproduced in previous fits to BEC DM \cite{schunck1998,bohmer07}. In the thin disk spiral NGC3198 our model only gives a small improvement over previous studies, and square and triangular results are practically the same. There are some discrepancies 
at large radii, which would likely be reconciled by a better description of the gas component.

The BEC DM model can accomodate data on different spiral types, 
and the parameters used are very similar. The vortex cores are 5.0-6.5 kpc in size with a lattice spacing of 1-2 kpc. We note that experiments have shown that the core and vortex-vortex distance can be of similar magnitude \cite{abo2001} in trapped condensates, although the core size is smaller than the distance. The theoretical vortex density expected in uniform condensates
is given by 
$n_\textrm{v}=m\Omega/\pi\hbar=0.05(m/10^{-22}\textrm{eV})(\Omega/10^{-16}\textrm{s}^{-1})\,\textrm{kpc}^{-2}$. 
For the spirals considered here we get $n_{\textrm{v}}\sim 0.10-0.13$ kpc$^{-2}$. The total number of vortices is then 
$N=\pi R^2 n_{\textrm{v}}\sim30-1000$ for radii $R\sim 10-50$ kpc. For the square lattice this gives
about 45 vortices within a distance of 50 kpc using the size and distance quoted above, which is a reasonable halo size for the Sc spirals discussed. 
Our model parameters are thus not completely untenable. With smaller cores we could get larger densities or have larger distance between vortices. However, we have found that in order to obtain the wiggles in the rotation curves the core size has to be roughly equal to the length 
between maxima and minima in the observations. In NGC3198 we have used a larger distance of 2.0 kpc to get a good fit, consistent 
with its rotational velocity being slightly lower than that of the others. Also for the dwarf NGC1560 we have used a smaller core size. This 
gives a larger vortex density which is consistent with a slightly larger rotational rate compared to the Sc samples. 

The wiggles in the rotation curves are located at short radii and studies have shown them to be correlated with the spiral arms \cite{beauvais1999,dicaire2008}, both facts strongly indicating baryonic processes as the origin. The current model implies that there could be 
additional effects in an ultralight BEC dark matter scenario. 
Assuming that the wiggles are caused by baryonic matter only, we can use the current model to extract
bounds on the parameters from the lack of vortex lattice effects in observations. This would in turn influence other models of BEC DM through limits on boson masses and two-body coupling terms. In particular, the absence of large wiggles produced by vortex lattices would
give a \emph{lower} limit on the two-body repulsion of the bosons. 

\section{Fine-Tuning Issues}\label{secfine}
The parameters of importance to the success of an ultralight BEC dark matter model are naturally 
the mass, which must be extremely small, and, for the self-interacting scenario, the strength of 
the two-body repulsion between the condensed particles. In Ref.~\cite{silverman02} BEC dark matter 
is suggested to arise from a single scaler field coupled to gravity undergoing spontaneous 
symmetry 
breaking to acquire a vacuum expectation value. This produces a 
cosmological constant $\Lambda$. The breaking of symmetry is done via a Ginzburg-Landau potential 
with quadratic and quartic terms. This gives mass and interaction terms to the scalar field. 
If we 
now make the additional assumption that the vacuum expectation value, $\phi_0$, arises from a 
mechanism that preserves parity (so that third-order terms can be ignored), the interaction term 
is
\begin{equation}
\frac{(mc^2)^2\phi^4}{4(\hbar c)^2\phi_{0}^{2}}=g\frac{\phi^4}{4},
\end{equation}
where we have restored constants of $c$ and $\hbar$. This terms is of 
course merely the standard interaction term in the Gross-Pitaevskii theory of interacting 
condensed bosons. We therefore see 
that the self-interacting scenario emerges from this procedure. 

The above scenario actually contains some additional information about the mass and interaction 
strength of the scalar field. One can derive a relation between mass, the interaction strength, 
and the cosmological constant. As in the previous sections, we want to use $g=4\pi\hbar^2a/m$ to 
express the interaction through the scattering length $a$. We have
\begin{equation}
a=\frac{m^5 G}{4\pi \hbar^6 \Lambda}=7\cdot 10^{-30} m_{22}^{5} 
\frac{\Lambda_0}{\Lambda}\,\text{fm},
\label{lamb}
\end{equation}
where $\Lambda_0=0.7\rho_{crit}\sim10^{-52}$ m$^{-2}$ is the best estimate of the 
cosmological constant from WMAP \cite{WMAP}. As is clearly seen, the value of $a$ in this model 
is quite far from the considerations of the previous sections, where 
$a\sim10^{-61}$ fm was found to be a favorable value. A mass of order 
$m\sim10^{-28}$ eV would be required to get this latter value of $a$ (this is still within the 
bounds on ultralight dark matter discussed in Ref.~\cite{car98}).

These considerations imply that the dark matter vortex lattice proposed here is far 
from generic and requires finely tuned parameters to work. However, the model of spontaneous 
symmetry-breaking discussed in Ref.~\cite{silverman02} is also highly speculative. The cause of the 
observed dark energy is still unknown and it is therefore reasonable to consider
scenarios where the relation in Eq.~\eqref{lamb} is not valid. This can easily be done by 
considering 
more general potentials than the standard Ginzburg-Landau scalar potential. The 
breaking of symmetry could then be accomplished in a manner that allows other relations between 
$m$ and $a$. This could allow other ranges of the parameters and perhaps make way for dark 
matter 
BEC vortex lattices without extreme fine-tuning. For lack of a better understanding, the 
cosmological constant, $\Lambda$, could be considered an outside parameter, we want to address
dark matter scenarios that are not necessarily from unified dark 
matter-dark energy models.

\section{Relevant Experiments and Numerical Simulations}\label{secatom}
The present experimental situation in ultracold atomic gases also encourages the hope that one 
could possibly probe the consequences of bosonic dark matter and condensates in laboratory 
experiments. Many groups around the world can routinely produce condensates of bosons, degenerate 
fermi gases, and also interesting mixtures of the two \cite{bloch2008,giorgini2008}. Another 
feature is the extreme control experimentalists exercise over the interactions between the 
different atomic species. This is achieved through the use of Feshbach resonances that allows the 
tuning of the interactions strength over many orders of magnitude and also whether the atoms 
repel or attract each other.

A very interesting proposal is that one can tune the inter atomic dipole-dipole 
interaction to resemble the gravitational force \cite{odell2000}. By illuminating the atomic 
clouds with a careful arrangement of multiple laser beams, it was suggested how to 
eliminate the $1/r^3$ dependence of the dipole-dipole interaction and keep only the $1/r$ part in 
the near-zone. The force constant of the remaining term depends on the laser intensity and atomic 
polarizability. The resulting force is equivalent to the attraction between two opposite electric 
charges with $q\sim e/2000$, so compared to electromagnetism it is not strong. However, in 
comparison, the tiny magnitude of gravity means that this force can be much larger than the 
normal gravitational force based on the atomic masses. 

If this proposal is successfully implemented in 
experiment, one could therefore simulate gravitational forces between the gas particles of both 
fermi and boson species, and with 'effective' masses that are vastly different from the given 
atomic masses. This could very likely be a way to test the movement of particles during 
virialization and see how structures form. Comparison with observations and $N$-body simulations 
would then allow us to expand our knowledge of the evolution of structure in our Universe. Since 
rotation is also routinely applied to the ultracold gases \cite{bloch2008,fetter2009}, there would also be ways to test the 
scenario considered in this paper with the formation of vortex lattices. One could then imagine a 
mixture 
of clouds with bosonic particles as dark matter and then fermions to represent luminous matter. 
This would allow the experimental exploration of the influence of BEC dark matter on structures in 
the luminous component. The ability to tune self-interactions and cross species interaction could 
also allow us to test effects of dark matter-normal matter interactions.

Although these experiments are extremely hard, the advances of recent years leave us hope that 
one could reach experimental capabilities that can explore some of these systems in the not too 
distant future. In fact, within traditional condensed matter the ideas of simulations for
instance cosmological phenomena using experiments have been around for some \cite{zurek1996}, 
and this type of thinking naturally extends to ultracold atomic gases which are often 
seen as a simulator for condensed matter systems.

Alternatively, the dynamics can be simulation on a computer. Comparison of
theory and experiment on ultracold atomic gases has shown that the Bose-Einstein condensates 
can be well described by the non-linear Gross-Pitaevskii equation \cite{BECbook} which can be solved 
numerically with great precision (see \cite{muru2009} for details on state-of-the-art numerical
techniques and references to relevant work). The inclusion of attractive $1/r$ gravitational
type potentials have been considered recently \cite{cart2008} in both a variational approach
and in numerical simulations. Rotation should be possible to include in similar fashion. We 
speculate that to separate dark and luminous matter one can consider a mixtures of two
bosonic atoms that are rotated differently so as to allow for vortices in one but not the 
other component or simply a bose-fermi mixture as also mentioned above. 
Issues of a potential thermal non-condensed part of the dark matter BEC could also 
be addressed numerically.

A recent example of the success of the Gross-Pitaevskii approach is worth mentioning. The
atom $^{52}$Cr has a large magnetic dipole moment that can be aligned by externally applied
fields. This means that an attractive dipole-dipole force can be created that was predicted
to produce collapse of a $^{52}$Cr condensate. This has been experimentally observed and found
to be in agreement with predictions from the Gross-Pitaevskii equation including 
the long-range force \cite{lahaye2008}. This implies that long-range forces can be accomodated
by the Gross-Pitaevskii theory as well.

\section{Conclusions}\label{seccon}
We have consider some models of dark matter where the main component is a bosonic particle with a 
very small mass of order $m\sim10^{-22}$ eV. Naively, this gives a de Broglie wavelength that is 
of galactic proportion such that one can imagine the entire galactic dark matter halo being in the 
condensate ground-state. We considered previous proposal with such particles with and without 
repulsive self-interactions and found general consistency between these for certain ranges of the 
mass and interaction strength. Moreover, we considered the suggestion 
that superfluid BEC dark matter in rotation would likely also lead to vortices as seen in 
atomic BEC experiments.

In case of a repulsive self-interaction we argued that the vortex size should be determined 
locally by the coherence length of this repulsive interaction. This means that we have two scales 
in the problem: A galactic one, given by the de Broglie wavelength from the tiny mass, and a 
sub galactic one that is determined by the mass and the two-body interaction strength 
(characterized by the scattering length $a$). We 
explored the consequences of self-interactions on the virialization of gravitationally bound 
structures and found almost no effect for reasonable values of $m$ and $a$.

Under the 
assumption of dark matter being an ultralight BEC, the rotation of spiral galaxies would cause 
vortex lattices to form. We briefly addressed possible effects on the growth of perturbations in 
the linear regime, although this is probably not relevant since the rotation rate in the early 
Universe is very small. We then considered possible effects of 
sub galactic vortices in the dark matter on the rotation velocity curves of virialized galaxies 
with standard dark matter halo profiles. Here we found that one can actually get substructure in 
the rotation curves that resemble some observations, but that this requires large vortex core size 
and small vortex-vortex distances. The mass and interaction strength needed to realize this where 
found to be fine-tuned, but could possibly 
be accommodated in more general setups. 

The present investigations and simple numerical experiments point to an interesting effect from 
bosonic dark matter. However, to fully explore the influence of vortex lattice formation and the 
feed-back on structure formation in luminous matter one would need to consider an ultralight BEC 
dark matter component in large $N$-body simulations.

\acknowledgments
Discussions with H.~O.~U. Fynbo and S. Hannestad are highly appreciated. Thanks also to J.~P.~U. Fynbo
for reading the draft and suggesting valuable improvements.

\end{document}